\documentclass[aps,pre,twocolumn,notitlepage,superscriptaddress]{revtex4-2}

\usepackage{amsmath}
\usepackage{amssymb}
\usepackage{hyperref}
\usepackage{graphicx}
\usepackage[arrow, matrix, curve]{xy}
\usepackage{bbm}
\usepackage{bm}
\usepackage{yhmath}
\usepackage{color}
\usepackage{physics}
\usepackage{bbold}
\usepackage{cleveref}
\usepackage{import}
\usepackage{graphicx}
\usepackage{calc}
\usepackage{dsfont}
\graphicspath{{Figures/}}

\usepackage{color}

\usepackage{mathtools}
\usepackage{esint}
\usepackage{upgreek}

\renewcommand\vec[1]{\boldsymbol{\mathrm{#1}}}

\usepackage[usenames,dvipsnames]{xcolor}
\hypersetup{colorlinks=true, linkcolor=BrickRed, urlcolor=blue!50!black, citecolor=blue!50!black}

\usepackage[normalem]{ulem}

\makeatletter
\newcommand\hide@visible[1]{%
  \bgroup\fboxsep=.3ex\colorbox{Gray}{begin hide}%
  #1\colorbox{Gray}{end hide}\egroup%
}
\newcommand\hide@hidden[1]{%
  \bgroup\fboxsep=.3ex\colorbox{Gray}{hidden text}%
}
\newcommand\hide@invisible[1]{}
\newcommand\makevisible{\let\hide\hide@visible}
\newcommand\makehidden{\let\hide\hide@hidden}
\newcommand\makeinvisible{\let\hide\hide@invisible}
\makeatother
\makehidden

\begin{document}	

\noindent\small Phys. Rev. E 113, 045409 (2026). \\
\vfill 
	
	\title{  
Free chiral self-propelled robots compared to active Brownian circle swimmers}
	\author{Thomas Kiechl}
	\affiliation{Institut f\"ur Theoretische Physik, Universit\"at Innsbruck, Technikerstra{\ss}e, 21A, A-6020 Innsbruck, Austria}
  	\author{Amy Altshuler}
	\affiliation{The Raymond and Beverley School of Chemistry, Tel Aviv University, Tel Aviv 6997801, Israel}
	\author{Anton L\"uders}
        \email{anton.lueders@uibk.ac.at}
	\affiliation{Institut f\"ur Theoretische Physik, Universit\"at Innsbruck, Technikerstra{\ss}e, 21A, A-6020 Innsbruck, Austria}
 	\author{Yael Roichman}
        \email{roichman@tauex.tau.ac.il}
	\affiliation{The Raymond and Beverley School of Chemistry, Tel Aviv University, Tel Aviv 6997801, Israel}
	\author{Thomas Franosch}
        \email{thomas.franosch@uibk.ac.at}
	\affiliation{Institut f\"ur Theoretische Physik, Universit\"at Innsbruck, Technikerstra{\ss}e, 21A, A-6020 Innsbruck, Austria}

	\date{\today}
	
	\begin{abstract}

Macroscopic active matter systems, such as bristle bots, provide a compelling platform for investigating nonequilibrium dynamics at highly visible scales. To fully leverage their accessibility, accurate mathematical models are needed to corroborate experiments. In this work, we study the motion of a free chiral Hexbug (Nano Newton series) via video tracking and compare the results to theoretical predictions from overdamped Langevin equations for active Brownian circle swimmers (ABCs). We find good agreement between the Hexbug’s dynamics and ABC model predictions, particularly for the mean-squared displacement (MSD) and the intermediate scattering function (ISF). Deviations between the Hexbug data and the ABC model arise primarily in the short-time behavior of the real-space propagator, where translational noise is most evident. Our results generally support the use of models based on overdamped Langevin equations as a robust framework for describing Hexbug motion when the influence of translational noise is negligible. Moreover, they demonstrate the sensitivity of ISF- and propagator-based analyses in characterizing active systems. Our approach opens new avenues towards refining coarse-grained models and advancing the theoretical understanding of macroscopic active systems.

	\end{abstract}
	
	
	\maketitle	
	
	\section{\label{sec:intro} Introduction}

Active particles are agents or microswimmers that convert energy into directed motion, driving them far from equilibrium. This definition includes a broad spectrum of systems, ranging from flocks of birds~\cite{bech2016}, and swarms of insects~\cite{Attanasi2014}, to artificial systems such as Janus particles and micro-machines ~\cite{Howse2007, Jiang2010, Volpe2011, Kurzthaler2018}. In recent years, active particles have become a focal point of interdisciplinary research, as their constant energy flow drives fascinating collective behaviors, such as cohesive groups~\cite{Attanasi2014, Stengele2022, Levay2025}, and enables the development of smart devices and materials~\cite{Guo2020}. A wide range of studies have explored experiments, theoretical models, and simulations with applications spanning biology~\cite{Needleman2017, Lipowsky2005}, social transport~\cite{Helbing2001}, and statistical physics~\cite{Cates2012, Chaudhuri2014, Fodor2016, Falasco2016, Speck2016, Fodor2018, Caraglio2022, Kiechl2024, Zanovello2021, Engbring2023}. Additionally, potential applications in robotics~\cite{Cheang2014} as well as in biomedicine~\cite{Wang2012, Henkes2020, Ghosh2020} have been investigated, including targeted drug delivery~\cite{Erkoc2019, Cheang2014, Gu2024} or decontamination of water and soil~\cite{Vilela2017, Gao2014, Guo2020}.

One notable example of artificial active particles is the group of Hexbugs or bristle bots, which are macroscopic (centimeter-sized) vibrating robots with bristle-like feet that convert their vibrations into directed forward motion. In the case of the \textit{Hexbug Nano Newton series} studied in this work, the motion also exhibits an intrinsic chirality~\cite{Altshuler2024}. Due to their affordability, ease of tracking, and the absence of complex hydrodynamic interactions, Hexbugs have been recognized as excellent toy models for experimental studies of active matter systems~\cite{Ning2024}. As a result, they have been widely employed in studies of collective behavior and movement in complex environments~\cite{Baconnier2022, Boudet2021, Dauchot2019, Giomi2013, DiBari2022, Callegari2023, Boriskovsky2024, Molodtsov2025, HernandezLopez2024}. Their simplicity also makes them ideal for educational purposes and science communication, as demonstrated in recent works advocating their use for teaching active-matter concepts~\cite{Callegari2023}.

Although Hexbugs are increasingly used in experimental studies, an ongoing debate remains about the appropriate unifying framework for describing their motion from a coarse-grained modeling perspective. In the literature, different strategies have been employed, with some studies favoring deterministic models~\cite{Fortunic2017, Kim2020, Becker2014} while others rely on stochastic Langevin equations frequently used to describe active colloids at microscopic and mesoscopic scales~\cite{Ning2024}. Within Langevin-equation-based models, overdamped dynamics are commonly applied, either incorporating translational noise~\cite{Anand2025, Caprini2024} or omitting it~\cite{Callegari2023, Giomi2013}. On the other hand, inertial terms are sometimes included to account for the macroscopic nature of bristle bots~\cite{Loewen2020, Scholz2018, Ning2024}. Here, the specific choice of the mathematical model is typically based on empirical observations, with equations of motion adjusted on a case-by-case basis~\cite{Ning2024, Baconnier2022}.

A recent advance in modeling Hexbug dynamics involves the identification of self-alignment terms, which are increasingly recognized as essential components in the equations of motion for bristle bots~\cite{Dauchot2019, Baconnier2022, Carrillo2025}. These terms contribute significantly to the dynamics when inertial effects are non-negligible or when the system is subject to external potentials and confinement. In most works addressing such conditions, they are typically still incorporated into different versions of the coarse-grained Langevin-type models~\cite{Dauchot2019, Baconnier2022, Carrillo2025}, which then act as the foundation for the assumed equations of motion of the macroscopic bristle bots. Although the role and necessity of these self-alignment contributions have now been examined in detail~\cite{Dauchot2019, Carrillo2025}, the validity of assuming Langevin equations as the basis for the corresponding models remains notably underexplored.

A promising tool for assessing whether Langevin-type models are suitable for describing the equations of motion of Hexbugs is the intermediate scattering function (ISF) --- the characteristic function of the system’s propagator. The ISF has proven highly effective in classifying the dynamics of free active particles at the microscale~\cite{Kurzthaler2016, Kurzthaler2018, Kurzthaler2024}. Unlike traditional metrics such as the mean-squared displacement (MSD)~\cite{Ebbens2012, Dietrich2017, Ebbens2014, Brown2015, Brown2014}, which fail to distinguish between run-and-tumble particles (RTPs), active Ornstein-Uhlenbeck particles (AOUPs), and active Brownian particles (ABPs)~\cite{Ebbens2010, Martens2012}, the ISF reveals distinct signatures for different types of active motion~\cite{Kurzthaler2016, Kurzthaler2018, Kurzthaler2024}. This distinction arises because RTP, AOUP, and ABP models yield identical MSDs but significantly different ISFs. Additionally, analyzing the probability density profile of free Hexbugs may offer further insight into their dynamics, as this method has successfully demonstrated the importance of including self-alignment terms under external potentials~\cite{Dauchot2019, Carrillo2025}.

Building on these insights, the objective of this work is to explore the accuracy of using the overdamped Langevin equation-based model of an active Brownian circle swimmer (ABC)~\cite{KurzthalerCircle2017, Rusch2024, Ning2024} as the foundation for the equations of motion of a Hexbug. To do so, we compare experimental data of a free Hexbug, where we assume the self-alignment terms to be negligible due to the absence of external potentials and confinement, to theoretical predictions for three key observables from the ABC model. First, we examine the MSD, a quantity frequently used to extract system parameters from experimental data~\cite{Carrillo2025, Siebers2023}. Next, we apply the ISF for the first time to bristle bots to validate the overdamped Langevin equation-based models as the foundation for their equations of motion. Finally, we compare the displacement probability density of the free Hexbug's motion with approximated short-time predictions, derived from the ABC model within this work. Here, the probability density offers insight into the translational fluctuations, which is usually hard to resolve for macroscopic active particles with large self-propulsion to noise ratios.

Our comparison between free chiral Hexbugs and the theoretical results~\cite{KurzthalerCircle2017} demonstrates that the ABC model generally provides a good qualitative description of the bristle bot's dynamics. In particular, the predicted MSD and the intricate features of the ISF are successfully reflected in the Hexbug's behavior. Only when analyzing the short-time probability density, differences between the Hexbug and the ABC emerge, which correspond to the regime where the translational noise is purely responsible for the variance of the displacement distribution.

This work is structured as follows: First, we briefly discuss the experimental setup, which is used to obtain the Hexbug measurements, as well as the equations of motion of an ABC in Sec.~\ref{sec:exp}. Following that, we present the results of the comparisons regarding the MSD, the ISF, and the propagator in Sec.~\ref{sec:results}. Finally, we conclude with an outlook in Sec.~\ref{sec:discussion}.

\section{Experimental Setup and Model} \label{sec:exp}

To compare the Hexbug dynamics to the predictions of the ABC model, we extract particle trajectories from video recordings and study if they align with analytic predictions. The following subsection provides details on the experimental setup used to obtain the Hexbug trajectories and the general mathematical framework describing ABCs.

\subsection{Experiments} \label{sec:experiments}

The experimental setup used to measure Hexbug trajectories follows the protocol established in previous studies by some of the present authors and collaborators~\cite{Altshuler2024, Vatash2025}. Hence, detailed information can be found in the corresponding publications. For completeness, a brief summary is given below.

\begin{figure}[ht!]
    \centering
    \includegraphics[width=\linewidth, trim=0 0 0 0, clip]{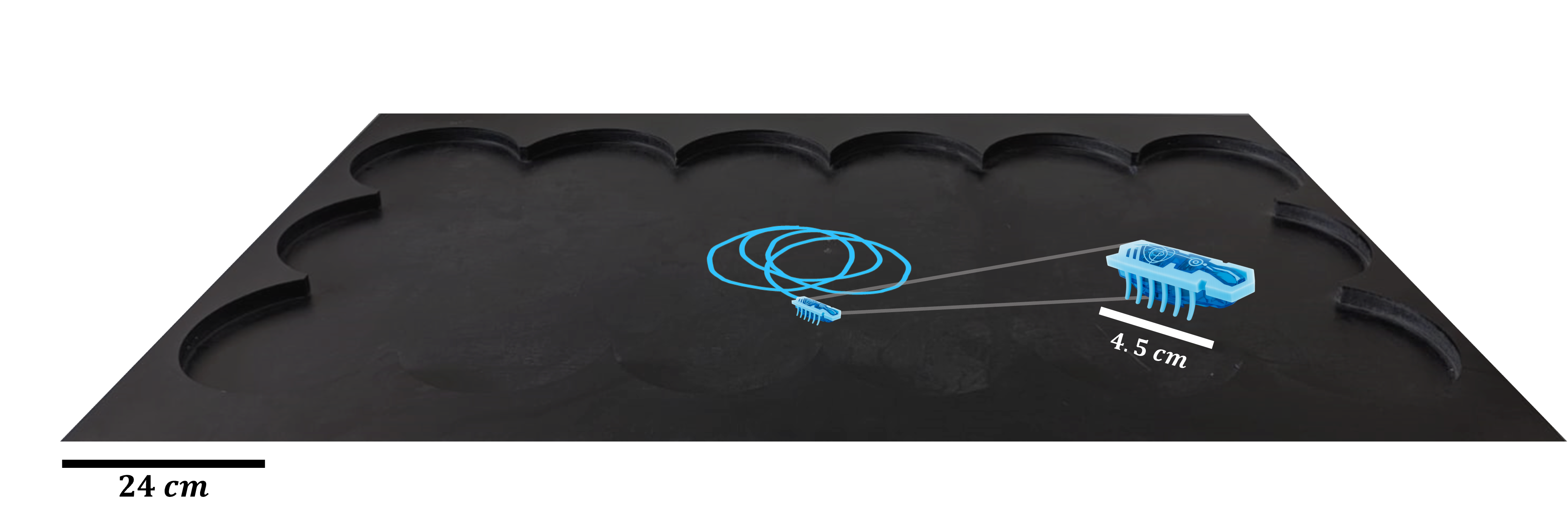}
    \caption{Arena utilized for the measurements of the Hexbug trajectories and its dimensions indicated by scale bars.}
    \label{fig:arena}
\end{figure}

A self-propelled, chiral bristle robot (Hexbug Nano Newton series) is placed in an arena with confining walls. The outer dimensions of the arena are $150 \times 90\, \textrm{cm}^2$, while the inner dimensions are $137 \times 63\, \textrm{cm}^2 $. The surface in direct contact with the bristle bots is made of medium-density fibreboard (MDF). The distinct roughness of the MDF material ensures high friction between the Hexbug and the surface, establishing overdamped particle dynamics. The chirality of the Hexbug, which consistently exhibited a counter-clockwise bias, was verified through measurements beforehand (see also the supporting information of Ref.~\cite{Altshuler2024}). Whenever a Hexbug touches the walls of the arena, it is manually repositioned at the center of the arena. The experimental setup is depicted in Fig.~\ref{fig:arena}.

\begin{figure}
    \centering
    \includegraphics[width=\linewidth, trim=0 0 0 0, clip]{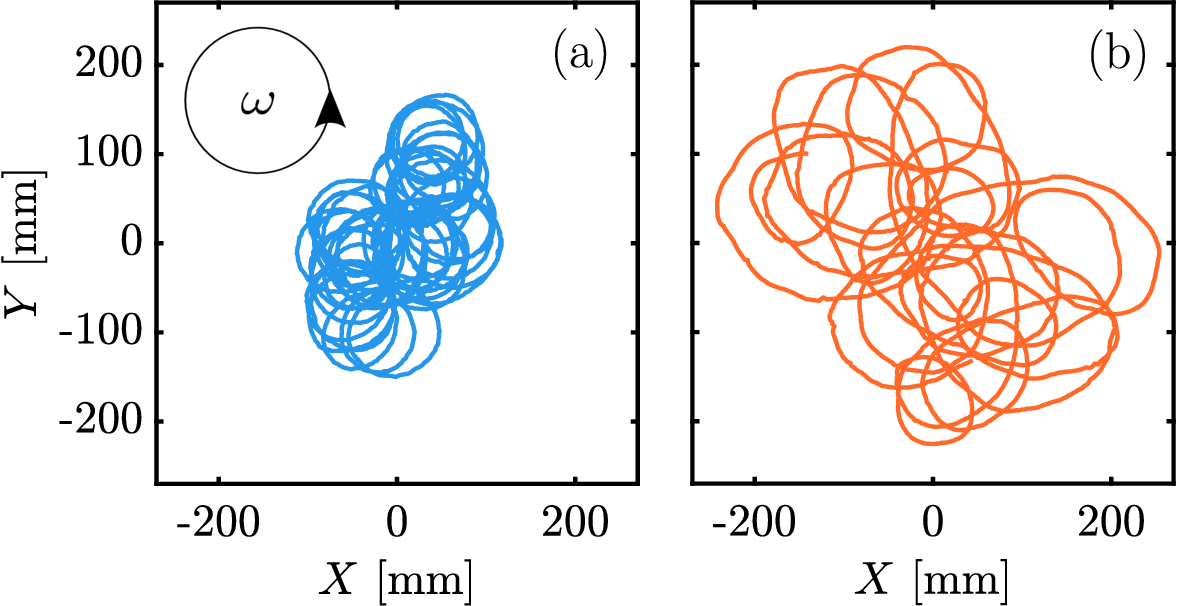}
    \caption{Trajectory segments of the two Hexbugs analyzed in this work, highlighting the strong variation in the particle velocity, chirality, and diffusivity of different units. The data corresponds to a time interval of approximately $ 50\, \textrm{s} $. The first Hexbug is dominated by the rotational bias (a). The second Hexbug is more diffusive (b).}
    \label{fig:trajectory}
\end{figure}

As also noted elsewhere~\cite{Siebers2023, Carrillo2025}, we find that the velocity and the chirality can deviate significantly between different Hexbug units of the same type and same manufacturer. To explore the accuracy and universality of the ABC model, we therefore analyze two bristle bots with vastly different dynamical properties. Example trajectories of the two chosen Hexbugs are presented in Fig.~\ref{fig:trajectory}~a) and b), highlighting the different dynamical behavior. Consequently, all presented averages of experimental data are calculated by sampling over a fixed Hexbug unit. To further ensure consistency, the batteries of the Hexbugs are regularly replaced during the experiments. 

The bristle bot trajectories are recorded using a Logitech BRIO 4K webcam at 60 frames per second (FPS). Conventional particle-tracking algorithms~\cite{Crocker1996} are employed to extract the trajectories. The positions are recorded as $(X,Y)$ coordinates in millimeters, with time represented as integer frame numbers. Due to USB bandwidth limitations during the real-time transfer of 60 FPS video from the camera to the computer, the recorded particle positions occasionally remain ``frozen'' or are only partially updated. To correct for this artifact, we process each recorded trajectory by filtering out any states where the displacement satisfies $|\Delta \vec{R}|\leq 0.05 \, \textrm{mm}$, where the mean step size between two frames is expected to be roughly of the order of $ 0.30\, \textrm{mm} $. 

Before analyzing the data, the first and last eighths of each Hexbug trajectory are removed to eliminate boundary wall interactions and potential influences from manually placing the Hexbug at the start of a measurement. The remaining cropped trajectories are then concatenated to form longer time series. This approach compensates for the finite size of the desk and ensures sufficient trajectory length to make meaningful statements about the diffusive long-time regime.

To concatenate the cropped trajectories into a single time series, we align both the orientation and position of the trajectory segments. In detail, to append a new trajectory segment, we first rotate it so that the direction defined by its first two positions matches the direction defined by the vector connecting the last two positions of the preceding trajectory. After rotation, the new segment is translated such that its starting point coincides with the second-to-last point of the preceding trajectory segment.

The concatenation of shorter trajectories is justified by the assumption that particle dynamics are approximately Markovian. To verify that this procedure does not significantly affect the results, we also analyzed the unmodified (unconcatenated) trajectories of the Hexbug with stronger rotational bias [Fig.~\ref{fig:trajectory}~a)]. The qualitative behavior was unchanged, and only minor quantitative variations appeared in parameters extracted from MSD fits, which were unweighted in the case of the unconcatenated trajectories ($v$: $3.7\%$ and $\omega$: $6.5\%$, see Sec.~\ref{sec:model}). As expected, larger deviations of $78.9\% $ and $74.7\%$ were observed in the rotational and translational diffusion coefficients $D_{\textrm{rot}}$ and $ D $, due to the inability of short trajectories to fully capture the long-time diffusive regime.

Additionally, we also tested whether the order of trajectory segments during concatenation influenced the results. The MSD fit parameters remained stable across different sequence arrangements, with negligible deviations: ($v$: $0.8\%$, $\omega$: $0.1\%$, and $ D_{\textrm{rot}} $: $ 2.4\%$ for the diffusive Hexbug). A larger deviation of $55.1\%$ was found for the translational diffusion coefficient. However, as discussed below (see Sec.~\ref{sec:msd}), this parameter is sensitive to small changes in the fit procedure as it cannot be reliably obtained from the MSD data.

\subsection{Equations of motion of an active Brownian circle swimmer (ABC)}\label{sec:model}

To explore the validity of Langevin-type models, we adopt the simplest set of Langevin equations that capture the necessary properties to reproduce the observed Hexbug trajectories (self-propulsion, chirality, and stochasticity) while possessing known analytic solutions for the corresponding MSD and ISF: the active Brownian circle swimmer (ABC). This approach is consistent with previous studies, such as Refs.~\cite{Callegari2023, Ning2024, Carrillo2025}, which modeled chiral bristle bots using similar frameworks. The applicability of the overdamped particle dynamics is further supported by our experimental results, as discussed below. Since we are only concerned with free particles, self-alignment terms are assumed to be negligible~\cite{Dauchot2019, Carrillo2025}.

In detail, the equations of motion for an ABC describe an overdamped particle undergoing translational diffusion with a diffusion constant $ D $, superimposed with directed motion at a constant speed $ v $ along its instantaneous orientation $\vec{u}(t) = (\cos\vartheta(t), \sin\vartheta(t))^T$. The corresponding orientational angle $ \vartheta $ evolves through rotational diffusion with a rotational diffusion coefficient  \( D_{\textrm{rot}} \), combined with a deterministic constant angular drift $ \omega $. The latter introduces a persistent rotational bias, modeling the intrinsic chirality observed in the Hexbug dynamics. The equations of motion are, thus, given by
\begin{subequations}
\begin{align}
    \dot{\vec{R}} &= v \vec{u} +  \sqrt{2D}   \vec{\eta}(t), \label{eq:EOM1}\\
    \dot{\vartheta} &= \omega +  \sqrt{2D_{\textrm{rot}}}   {\xi}(t), \label{eq:EOM2}
\end{align}
\end{subequations}
where the scalar noise \( \xi(t) \) and the components of the noise vector \( \vec{\eta}(t) \) are independent random variables, each following a Gaussian distribution with zero mean and delta-correlated variance, i.\@e.\@, $\langle \xi(t) \xi(t') \rangle = \delta(t - t')$, and $ \langle \vec{\eta}(t) \otimes \vec{\eta}(t')  \rangle = \mathds{1} \delta(t-t') $. Here, $ \mathds{1} $ is the unit matrix, and $ \otimes $ is the dyadic product.

\section{Results} 
\label{sec:results}

The dynamics of the free Hexbug is examined in detail by calculating the MSD, the ISF, and the propagator using the displacements from the Hexbug trajectory, which limits the time resolution to the frame rate of the video tracking. The theoretical predictions for the MSD and ISF of an ABC are taken from Refs.~\cite{KurzthalerCircle2017, Loewen2020}, which provides a detailed study of such particles. An approximate analytical form for the propagator is derived as part of this work. For maximum comparability with other macroscopic active agents and theoretical models, the length scale and time scale are always presented in physical units of $ \textrm{mm} $ and $ \textrm{s} $, as well as in units of $ v/\omega $ and $ 1 / \omega $.

\subsection{Mean-square displacement (MSD)}\label{sec:msd}

The MSD is widely used for the initial characterization and classification of particle dynamics in both Brownian and active matter systems, owing to its ease of computation across experiments, simulations, and theoretical models. Accordingly, it serves as the starting point for our analysis. Formally, the MSD is defined as
\begin{align}
    \langle \Delta \vec{R}^2 (t) \rangle= \left\langle \left| \vec{R}(t+\tau) - \vec{R}(\tau) \right|^2 \right\rangle, \label{eq:MSDDEF}  
\end{align}
where $\left\langle \cdot\right\rangle$ denotes the average over all displacements of the given lag time $t$. For an ABC (i.\@e.\@, Eqs.\ref{eq:EOM1} and \ref{eq:EOM2}), the MSD evaluates to~\cite{Loewen2020, KurzthalerCircle2017, Teeffelen2008, Ning2024}
\begin{align}
&\langle \Delta \vec{R}^2(t) \rangle = 4Dt\notag \\ &+\frac{2 v^2 }{(\omega^2 + D_{\textrm{rot}}^2)^2 }
 \Big[ \omega^2 - D_{\textrm{rot}}^2 + D_{\textrm{rot}}(\omega^2 + D_{\textrm{rot}}^2)t \notag\\
 &- e^{-D_{\textrm{rot}} t} \big[(\omega^2 - D_{\textrm{rot}}^2)\cos(\omega t) + 2D_{\textrm{rot}}\omega \sin(\omega t)\big] \Big].
 \label{eq:MSDTHEO}
\end{align}

\begin{figure}
    \centering
    \includegraphics[width=3.375in]{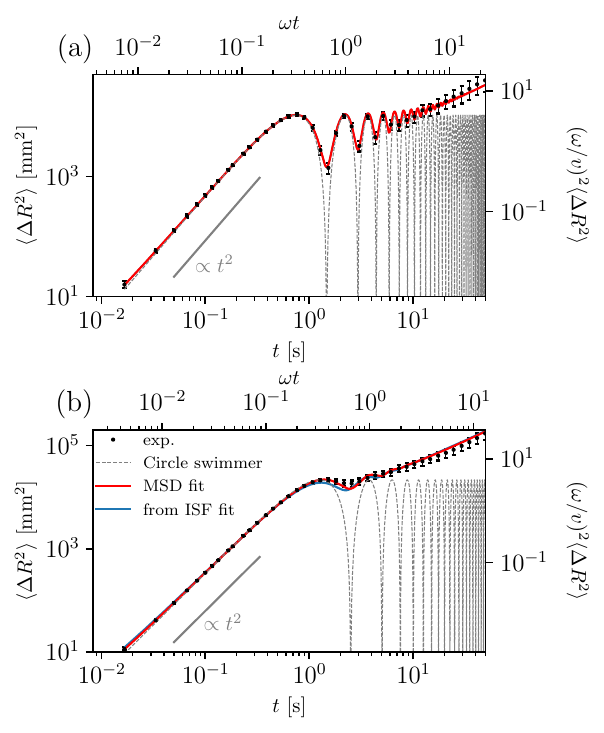}
    \caption{Mean-squared displacement (MSD) of the experimental data (black) compared to the analytical solution (continuous line) for an active circle swimmer (ABC) with MSD fit parameters (red) and ISF fit parameters (blue). The gray dashed line corresponds to a perfect circle swimmer (calculated with the MSD fit parameters). (a) Hexbug with a dominant rotational bias. (b) Hexbug with a more dominant rotational diffusion.
    }
\label{fig:msd_fit}
\end{figure}

To compute the MSD from the experimental data, we can characterize the Hexbug trajectory corresponding to the time $ T = N\Delta t  $ via $R(T) = (\vec{R}_0, \vec{R}_1, \dots, \vec{R}_N)$, with $\vec{R}_i = \vec{R}(i\Delta t)$. Here, $R(T)$ is a time-ordered set consisting of a total number of $N+1$ positions with the time ${\Delta t = 0.0167\ \textrm{s}}$ between two consecutive positions $ \vec{R}_k $ and $ \vec{R}_{k+1} $. From $R(T)$, we evaluate Eq.~\eqref{eq:MSDDEF} by averaging over all $ N_{\textrm{sample}}(t) $ possible displacement segments $ \vec{R}_k - \vec{R}_l $ for a given lag time $ t =(k-l) \Delta t \in \{0, \Delta t, 2\Delta t ,\dots,N\Delta t\}$. This approach is a standard strategy for obtaining accurate MSD data from time-averaging. The experimental results for the two studied Hexbugs are shown in Fig.~\ref{fig:msd_fit}~a) and b) via the symbols. The depicted error bars, which are given to indicate the weights used for the curve fitting, are conservative approximations for the standard deviation of the mean value adjusted for the correlations of the overlapping trajectory pieces and the persistence of the active motion. Detailed for the corresponding estimations are provided in Appendix~\ref{app:ERROR}.

The solid red lines in Figs.~\ref{fig:msd_fit}~a) and b) represent fits based on the theoretical prediction Eq.~\eqref{eq:MSDTHEO}, where the approximated error bars were used as weights during the fitting process. For comparison, the dashed gray line shows the MSD for a particle moving on a perfect circular orbit (i.e. a circle swimmer without translational or rotational noise)~\cite{KurzthalerCircle2017}
\begin{equation}
   \langle \Delta \vec{R}^2(t) \rangle = 4 \tilde{R}^2 \sin^2(\omega t/2), \label{eq:SW}
\end{equation}
which is included to highlight the diffusive characteristics of the experimental MSD data. The radius $ \tilde{R}=v/\omega $ and the angular frequency $ \omega $ are taken from the fit parameters. For both fitting and visualization in Fig.~\ref{fig:msd_fit}, not all possible lag times were included. Instead, a subset of lag times was selected, evenly spaced on a logarithmic scale, to avoid oversampling at longer $ t $ and overcrowded figures. For the Hexbug with the larger rotational bias, the MSD curve is fitted between $0.0167\ \textrm{s}$ and $ 100 \ \textrm{s} $. For the more diffusive bristle bot, the fit is obtained in the interval $0.0167\ \textrm{s} < t < 16.67\ \textrm{s}$. 

As highlighted by the initial persistent regime, which exhibits a $t^2$ dependence (see Fig.~\ref{fig:msd_fit}), the MSD data does not resolve the translational diffusion with high precision. While the parameters $v, \omega,$ and $D_{\text{rot}}$, as well as the general qualitative behavior of the resulting curves, are robust against variations of the fit interval (yielding deviations on the order of only a few percent), the translational diffusion coefficient $D$ is significantly more sensitive and can vary by a factor of two or three. To establish a physically consistent value for $D$, we optimized the fit intervals to align with our analysis of the propagator width at $\omega t = 1$ (see below).

While showing completely different characteristics, the MSD curves of both analyzed Hexbug units match the expected progression of an ABC~\cite{KurzthalerCircle2017} well (in particular for the bristle bot with the larger rotational bias). The resulting fit parameters are $v = (21.81 \pm 0.05) \ \mathrm{cm/s},\ 
\omega=(4.258 \pm 0.007)\ \mathrm{s^{-1}},\ 
D_{\textrm{rot}} = (0.082 \pm 0.005) \ \mathrm{s^{-1}},\ 
D = (0.35 \pm 0.07)  \ \mathrm{cm^2/s}$ for the data shown in Fig.~\ref{fig:msd_fit}~a) and $v = (18.31 \pm 0.04)\ \mathrm{cm/s},\ 
\omega=(2.497 \pm 0.009)\ \mathrm{s^{-1}},\ 
D_{\textrm{rot}} = (0.321 \pm 0.005) \ \mathrm{s^{-1}},\ 
D = (0.29 \pm 0.05)  \ \mathrm{cm^2/s}$ for the data shown in Fig.~\ref{fig:msd_fit}~b). The given uncertainties are the $68\%$ intervals of the fit.

To gauge the goodness of fit, we computed the reduced chi-squared statistic, $\chi_{\mathrm{red}}^2$. We obtain $\chi_{\mathrm{red}}^2 = 0.12$ for Fig.~\ref{fig:msd_fit}a and $\chi_{\mathrm{red}}^2 = 0.33$ for Fig.~\ref{fig:msd_fit}b. These notably low values reflect the conservative nature of our uncertainty estimates and the correlations between MSD points at different lag times (Appendix~\ref{app:ERROR}); as such, $\chi_{\mathrm{red}}^2$ here should be viewed as a relative diagnostic rather than a calibrated goodness-of-fit metric. Nonetheless, the low values are consistent with the good agreement between the theoretical prediction and the Hexbug data prior to the onset of the diffusive regime (visible by inspection of Fig.~\ref{fig:msd_fit}). This agreement is in line with previous studies~\cite{Ning2024, Callegari2023, Carrillo2025}, which successfully characterized Hexbugs using an MSD analysis.

Notably, the data presented in Fig.~\ref{fig:msd_fit}~a) exhibits the oscillating signature, which is characteristic of ABCs with dominant rotational bias. The calculated fit perfectly reproduces these features. The motion of the second presented Hexbug [Fig.~\ref{fig:msd_fit}~b)] is more diffusive, and consequently, this feature is less pronounced. Altogether, our analysis demonstrates that the ABC model provides an excellent quantitative description of the Hexbug’s MSD, capturing both its diffusive and oscillatory features with high fidelity.

\subsection{Intermediate Scattering Function (ISF)}\label{sec:isf}

As described in the introduction, it is well-established from studies of active particles on the microscale that the MSD alone can be insufficient to fully characterize different types of propulsion~\cite{Dauchot2019, Ebbens2010, Martens2012, Kurzthaler2018}. An alternative quantity that has proven effective in resolving subtle differences between distinct classes of active motion is the ISF~\cite{Kurzthaler2016, Kurzthaler2018, Kurzthaler2024}. The ISF can be calculated directly from the measured trajectory using its definition
\begin{align}\label{eq:isf_def}
    F(\vec{k}, t) &=  \left\langle \exp \bigl[-i \vec{k} \cdot \Delta \vec{R}(t)\bigr] \right\rangle,
\end{align}
and it acts as the moment-generating function of the particle displacements, providing both spatial and temporal information on the system.

By definition of the average in Eq.~\eqref{eq:isf_def}, the ISF is the spatial Fourier transform of the propagator, or probability density $ \mathbb{P}(\vec{r},\vartheta,t\mid\vartheta_0)$, averaged over the initial angle $ \vartheta_0 $ and integrated over the final angle $ \vartheta $ (see Ref.~\cite{KurzthalerCircle2017}). To specify the propagator in more detail, $ {\mathbb{P} \equiv\mathbb{P}(\Delta \vec{r} = \vec{r} - \vec{r}_0,\vartheta,t\mid\vartheta_0)} $ is the conditional probability density of being at position $\vec{r}$ with orientation $\vartheta$ after a lag time $ t $, given the initial position $\vec{r}_0$ and initial orientation $\vartheta_0$ at initial time $t_0 = 0$. For an ABC, the propagator satisfies the Smoluchowski equation
\begin{align}\label{eq:Fokker-Planck}
    \partial_{t} \mathbb{P}  = 
    - v \vec{u}\cdot\nabla_{\vec{r}} \mathbb{P}
    - \omega \partial_{\vartheta} \mathbb{P}
    + D_{\textrm{rot}} \partial_{\vartheta}^{2} \mathbb{P} + D \nabla_{\vec{r}}^2 \mathbb{P}.
\end{align}
This relation can be derived from Eqs.~\eqref{eq:EOM1} and \eqref{eq:EOM2} using standard methods of stochastic calculus.

Detailed derivations for the ISF of an ABC can be found in Refs.~\cite{Rusch2024, KurzthalerCircle2017}. Taking the solution out of Ref.~\cite{KurzthalerCircle2017} (simplified for isotropic translational noise), the theoretical prediction reads
\begin{align} \label{eq:isf_analytic}
&F(k,t)= \nonumber\\
& \frac{e^{-D k^2t}}{4\pi^2} \sum_{n=-\infty}^{\infty} e^{-a_{2n} D_\text{rot} t/4} 
\left[ \int_{0}^{2\pi} \mathrm{d}\theta \, \text{ee}_{2n} (q,M,\theta/2) \right]^2. 
\end{align}
Here, $ k = | \vec{k} | $, the $a_{2n}$ are the eigenvalues corresponding to eigenvalue equation
\begin{equation}
    L z(x) = a z(x),
\end{equation}
of the non-Hermitian Sturm-Liouville operator
\begin{equation}
    L = L(q, M) = \frac{d^2}{dx^2} + 2q \cos(2x) +  4\pi M \frac{d}{dx},
\end{equation}
and $x = \theta/2$. The functions $ \text{ee}_{2n}(q, M, x) $ are deformations of the complex exponentials $\exp(2nix)$ given by
\begin{equation}
    \text{ee}_{2n}(q, M, x) = \sum_{m=-\infty}^{\infty} A_{2n}^{2m}(q,M) e^{2 i m x},
\end{equation}
where $A_{2n}^{2m}(q,M)$ denotes the $2m$-th Fourier coefficient of $\text{ee}_{2n}(q, M, x)$ and the direction of the wave vector $ \vec{k} $ is chosen parallel to the $x $-axis~\cite{KurzthalerCircle2017}. Note that the solution for the ISF depends on the dimensionless quality factor $ M = \omega / 2\pi D_{\text{rot}} $, quantifying the deterministic circular motion with respect to the rotational diffusion, and the deformation parameter $q = 2ivk/D_\text{rot}$. In our case, we have approximately $ M = 8.26 $ and $ M = 1.24 $ for the Hexbug with a strong rotational bias [see Fig.~\ref{fig:trajectory}~a) and Fig.~\ref{fig:msd_fit}~a)] and the more diffusive Hexbug [see Fig.~\ref{fig:trajectory}~b) and Fig.~\ref{fig:msd_fit}~b)], respectively, when inserting the fit parameters obtained from the MSD curves.

\begin{figure*}
    \centering
    \includegraphics[width=1\linewidth, trim=0 0 0 0, clip]{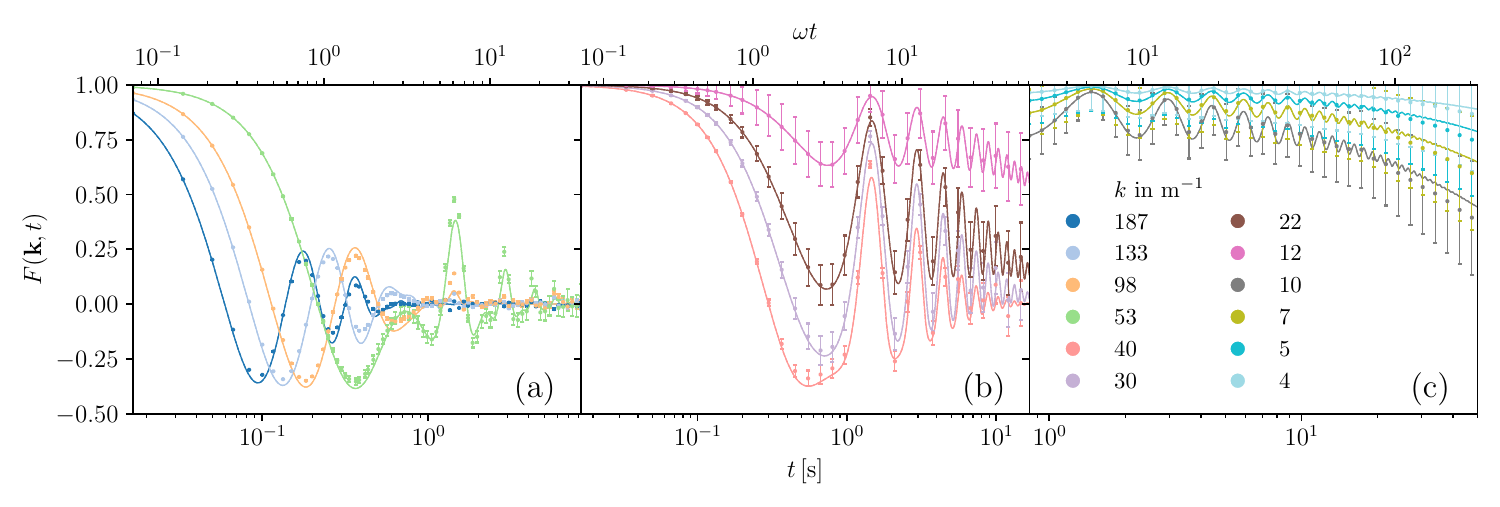}
    \vspace{0.5em} 
    \includegraphics[width=1\linewidth, trim=0 0 0 0, clip]{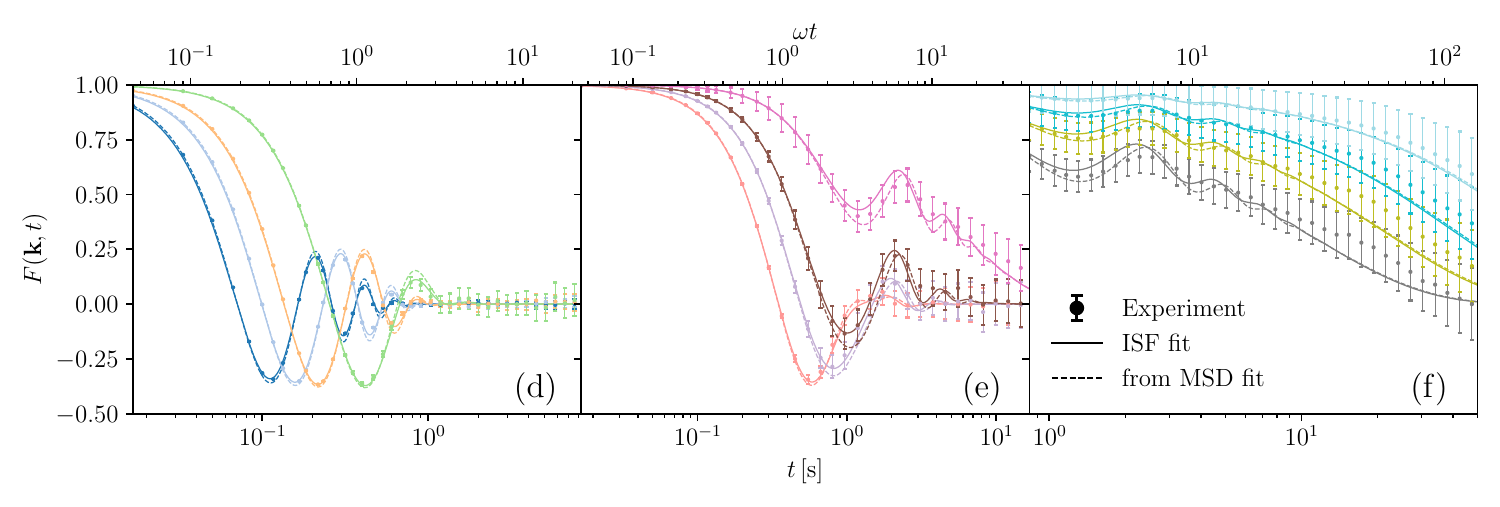}
    \caption{
    Intermediate scattering functions (ISFs) for various wavenumbers $k$ (different colors) from experiment (symbols) and fitting the theoretical results (continuous lines). Dashed lines are the theoretical predictions using the parameters from the MSD fit. Depicted are: ISF for strong rotational bias at large (a), intermediate (b), and small $ k $ (c). ISF for a more diffusive Hexbug at large (d), intermediate (e), and small $ k $ (f).}
    \label{fig:isf_comparison}
\end{figure*}

Figures~\ref{fig:isf_comparison}~a) to f) show the ISF for the two Hexbugs analyzed in this work. The symbols are the experimental results, where the error bars were approximated in the same way as described for the MSD in Appendix~\ref{app:ERROR} (using the same value for the effective sample size of independent trajectory pieces used to calculate the mean value of the ISF). The dashed lines correspond to the theoretical results using the fit parameters of the MSD analysis. The solid lines result from fitting the analytical expression for the ISF directly to the experimental data using a multidimensional fit based on the Broyden-Fletcher-Goldfarb-Shanno (BFGS) algorithm~\cite{Kurzthaler2024, Byrd1995}. In this approach, the ISF curves for $15$ different wavenumbers $ k $ were treated as independent components of a vector field and simultaneously fitted to the experimental data. For the curve fitting and for the plotting of the theoretical ISF, we use the representation of the ISF series expansion of Ref.~\cite{Rusch2024}. For the Hexbug with the stronger rotational bias, the fits are calculated for lag times in the interval $ 0.5\ \textrm{s} < t < 16.67\ \textrm{s} $, using data points equally spaced on a logarithmic scale. For the diffusive Hexbug, the fit is done for $ 0.167\ \textrm{s} < t < 15.0 \ \textrm{s} $. As initial guess for the parameters used as input for the BFGS algorithm, we use the values obtained by the MSD fit.

For the Hexbug with the stronger rotational bias
[Figs.~\ref{fig:isf_comparison}~a), b), and c)], the ISF fits converge to the same set of parameters as obtained from the MSD fit. In this case, the parameters extracted from the MSD already represent a robust minimum of the ISF cost function (i.\@e.\@, the weighted sum of squared residuals). The pronounced oscillatory behavior of the MSD contains sufficient information to accurately determine the system parameters from $\langle \Delta \vec{R}^2(t) \rangle$ alone. Notably, the corresponding ISF clearly exhibits features such as oscillations around finite plateaus and non-monotonic decay, which were identified as characteristic fingerprints of ABC dynamics in Ref.~\cite{KurzthalerCircle2017}.

For the more diffusive Hexbug [Figs.~\ref{fig:isf_comparison}d), e), and f)], the MSD lacks similarly sensitive features, which leads to slightly different results when fitting the data using the MSD or the ISF. In this case, the dedicated ISF fit yields a slightly better match to the oscillations of the experimental data at large wavenumbers $k$. For small wavenumbers, it is not decisive which parameter set provides the better agreement. The resulting parameters are 
$v = (18.401 \pm 0.002)\,\mathrm{cm/s}$, $\omega = (2.702 \pm 0.004)\,\mathrm{s^{-1}}$, $D_{\textrm{rot}} = (0.381 \pm 0.002)\,\mathrm{s^{-1}}$, and $D = (0.4309 \pm 0.0007)\,\mathrm{cm^2/s}$. The given uncertainties are again the $68\%$ intervals of the fit. The blue line in Fig.~\ref{fig:msd_fit}b) corresponds to the theoretical 
prediction of the MSD using the parameters obtained from the ISF fit. The resulting curve also aligns well with the experimental MSD data, further supporting the consistency of the extracted parameters and the robustness of the ISF fit.

The reduced chi-square values for the ISF fits are 
$\chi_{\mathrm{red}}^2 = 3.95$ for the Hexbug with the stronger rotational bias and $\chi_{\mathrm{red}}^2 = 0.38$ for the more diffusive bristle bot. The latter again reflects the conservative nature of our uncertainty estimates (see Appendix~\ref{app:ERROR} and note again that, due to correlations and heuristic weights, $ \chi^2_{\textrm{red}} $ is a relative diagnostic rather than a calibrated ``goodness-of-fit'' metric). The larger value for the Hexbug with the stronger rotational bias originates primarily from the curves at large wavenumbers: although the theoretical predictions capture the intricate features of the ISF, [see Fig.~\ref{fig:isf_comparison}~a)], they fall outside the comparatively small error bars in this regime. Given the macroscopic nature of the bristle bots, such minor deviations at high spatial resolution are unsurprising (see also our discussion of the propagator) and do not detract from the overall physical agreement.

The theoretical predictions of the ISF for the ABC model match the experimental data for both analyzed bristle bots. Given that the ISF exhibits intricate structure and nontrivial features across a wide range of wavenumbers $k$, this agreement provides strong evidence that modeling chiral Hexbugs as ABCs is well justified.

\subsection{Propagator}\label{sec:propagator}

Both the MSD and the ISF have their most characterizing features starting at intermediate timescales, making an assessment of small displacements and translational noise more difficult. This is, for instance, also highlighted by the lack of an initial diffusive regime in the MSD data shown in Fig.~\ref{fig:msd_fit}, which directly starts with a $ t^2 $ increase. A suitable candidate for resolving the characteristics of the translational fluctuations is the angle-independent propagator $ \mathbb{P}(\vec{r},t) $, which we define as
\begin{align}
   \mathbb{P} (\vec{r},t) = \int_{-\infty}^{\infty} \textrm{d} \vartheta  \int_{0}^{2\pi}   \frac{\textrm{d} \vartheta_0}{2\pi} \; \mathbb{P}(\vec{r}, \vartheta, t \mid\vartheta_0), \label{eq:PropDefine}
\end{align}
where $ \mathbb{P}(\vec{r}, \vartheta, t \mid\vartheta_0) $ is the full propagator, which we introduced earlier. Hence, this angle-independent propagator is obtained by averaging over the initial orientation $ \vartheta_0 $ and marginalizing over all possible final orientations $ \vartheta $. It provides a more detailed description of the system's dynamics by characterizing the probability density of particle displacements over time. This makes it a powerful tool for detecting subtle deviations and further analyzing the accuracy of the ABC model. Also, it gives direct access to the translational diffusion coefficient, which is hard to obtain by using only the other two observables.

Before turning to the experimental data, we first derive an analytical approximation for $\mathbb{P}(\vec{r},t) $ of an ABC in the short-time regime. When $t \ll 1/D_{\textrm{rot}} $ and $ M \gg 1 $, the MSD curve of the Hexbug, representing the second moment of its displacement probability density, resembles that of a particle moving along a perfect circular orbit (see Fig.~\ref{fig:msd_fit}, where the MSD curves initially follow the dashed curves corresponding to the perfect circle swimmer). This observation motivates approximating the ABC’s rotational short-time dynamics using the deterministic equation
\begin{align}
    \vartheta(t) & = \vartheta_0 + \omega t,
\end{align}
for a circle swimmer without rotational noise. The deterministic part of the translational motion then evaluates to
\begin{align}
        \vec{R}_c( \vartheta_0, t) =  \tilde{R} \,
        \textsf{D}(\vartheta_0) \, \begin{pmatrix}
            \sin(\omega t) \\ 
            1-\cos(\omega t)
        \end{pmatrix},
\end{align}
where $\tilde{R}=v/\omega$ is the radius of the circular orbit, and $ \textsf{D}(\vartheta_0) $ is the rotation matrix
\begin{align}
    \textsf{D}(\vartheta_0) =
    \begin{pmatrix}
    \cos \vartheta_0 & -\sin \vartheta_0 \\
    \sin \vartheta_0 & \cos \vartheta_0
    \end{pmatrix}.
\end{align}
For simplicity, we assume $ \vec{r}_0 = 0 $ and $ t_0 = 0 $ as the initial conditions.

In this zero-rotational-noise approximation, only translational diffusion causes the deterministic trajectory to spread over time. As a result, the system can be approximated by a Gaussian distribution centered around the deterministic trajectory $ \vec{R}_c(\vartheta_0, t) $. Thus, the full propagator can be expressed as
\begin{align}
       &\mathbb{P}(\vec{r}, \vartheta, t \mid \vartheta_0) \nonumber \\ &\approx \frac{1}{4\pi Dt} \exp\!\left(-\frac{|\vec{r} -  \vec{R}_c( \vartheta_0, t)|^2}{4D t}\right) \, \delta(\vartheta -  \vartheta(t)),
\end{align}
where the delta function enforces the deterministic evolution of the orientation $ \vartheta(t) $. Substituting this expression into Eq.~\eqref{eq:PropDefine} and performing the integration yields
\begin{align} \label{eq:p(r)}
        4 \pi & D t \; \mathbb{P}(\vec{r}, t) \nonumber \\ =&  \, \exp\!\left[-\frac{\vec{r}^2+4\tilde{R}^2\sin^2(\omega t / 2)}{4Dt}\right]\, I_0\!\left(\frac{| \vec{r} |\tilde{R}\sin(\omega t / 2)}{Dt}\right),
\end{align}
valid for $ t \ll  1/D_{\textrm{rot}} $ and $ M \gg 1 $. Here, $I_0(\cdot)$ is the modified Bessel function of the first kind to order zero.

Note that $ \mathbb{P}(\vec{r}, t) $ is invariant under a common rotation of $\vec{r}$, which means that it depends only on the magnitude of the displacement $ r = | \vec{r} | $. Thus, we can write $ {\mathbb{P}(\vec{r},t) = \mathbb{P}(r,t)} $. Moreover, since $\mathbb{P}(r,t) $ is a probability density, it must satisfy the normalization condition
\begin{equation}
    1 = \int \mathbb{P}(r,t) \, \textrm{d}^2 r = \int 2\pi r \, \mathbb{P}(r,t) \, \textrm{d} r ,
\end{equation}
for arbitrary times $ t $. Here, $ 2\pi r \, \mathbb{P}(r,t) \, \textrm{d} r $ is the probability for an ABC to be displaced by an absolute distance $ r $ during a lag time $ t $ from its initial condition.

To compute $ 2\pi r \, \mathbb{P}(r,t) $ from the experimental data, we construct a histogram with a bin size $ \Delta r_{\textrm{bin}} $ depending on the lag time $ t $. For each bin, we count all displacements $ r $ that fall within the intervals $ [n\Delta r_{\textrm{bin}}, (n+1)\Delta r_{\textrm{bin}}]$ with $n=0,1\dots,a $ and $ (a+1) \Delta r = r_{\textrm{max}}$. The resulting histogram is normalized such that its integral equals unity. Finally, we plot the normalized counts at the centers of the bin intervals.

\begin{figure*}[t] 
    \centering
    \includegraphics[width=\linewidth, trim=0 0 0 0, clip]{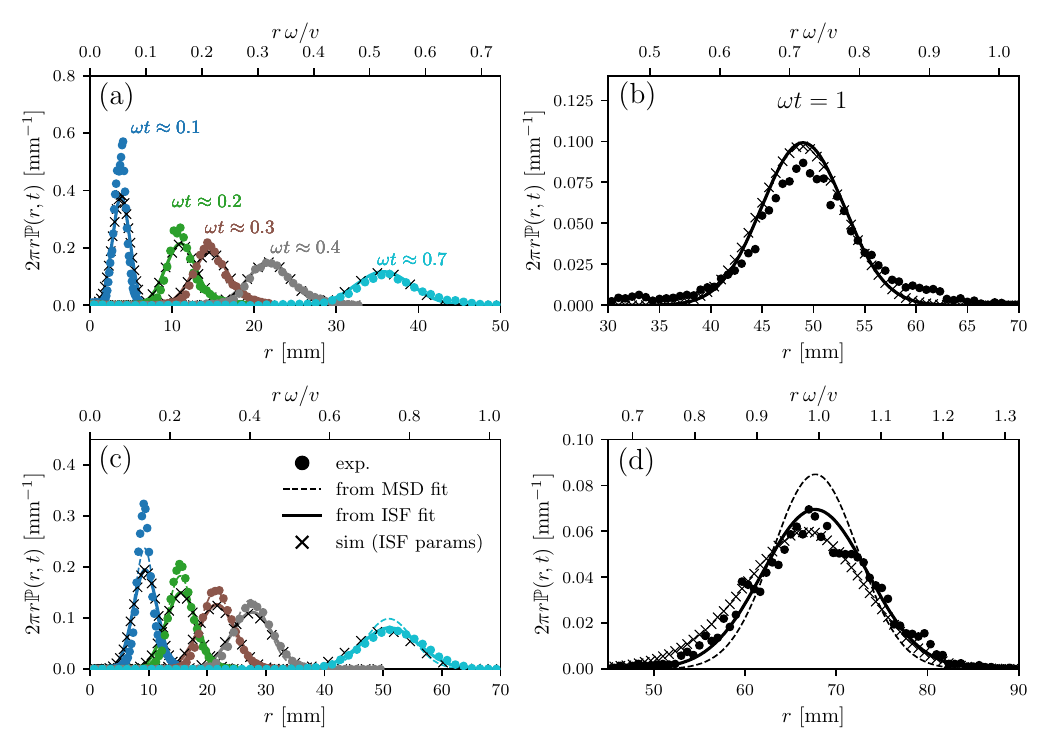}
    \caption{%
Probability distribution $2\pi r \, \mathbb{P}(r,t)$ of the displacement $r$ from the initial position at time $t$ for an active Brownian circle swimmer (ABC).  
Dashed lines: theoretical prediction using MSD parameters.  
Solid lines: theoretical prediction using ISF parameters.  
Circles: experimental Hexbug data.  
Crosses: simulations using ISF fit parameters.  
Panels (a) and (c) (short times): $\omega t = 0.1$ (blue), $0.2$ (green), $0.3$ (brown), $0.4$ (grey), and $0.7$ (light blue).  
Histograms are computed for displacements $r = |\vec{r}|$ using $50$ bins and maximum radius $r_{\mathrm{max}} = [6,20,22,33,50]$ mm for the Hexbug with the strong rotational bias, and $r_{\mathrm{max}} = [15,25,35,50,70]$ mm for the more diffusive Hexbug.  
Panels (b) and (d) (long times): $\omega t = 1$.  
Histograms are computed with $150$ bins and $r_{\mathrm{max}} = 100$ mm. Panels (a) and (b): experimental data for the Hexbug with the strong rotational bias. Panels (c) and (d): experimental data for the diffusive Hexbug.  
}
    \label{fig:propagator}
\end{figure*}

Figures~\ref{fig:propagator}~a) and b) display the experimentally measured $2\pi r \, \mathbb{P}(r,t) $ at short timescales and at $t \lesssim 1/\omega$ for the Hexbug exhibiting dominant rotational bias, respectively. Panels~\ref{fig:propagator}~c) and d) show the corresponding results for the more diffusive bristle bot. In each case, the solid line represents the analytical approximation of the ABC propagator, obtained using fit parameters extracted from the ISF. For the diffusive unit, the dashed line denotes the theoretical prediction calculated using the MSD fit parameters. 

To validate the accuracy of our analytic approximations, we compare the theoretical predictions with the propagator obtained by numerically integrating the ABC equation of motion using the stochastic Euler algorithm and the ISF parameters (crosses)~\cite{Git}. For the Hexbug, which exhibits a stronger rotational bias ($M = 8.26$), the simulations confirm that our approximations closely match the ABC propagator. In contrast, for the more diffusive unit with a weaker rotational bias ($M = 1.24$), the approximated propagator remains valid only at short lag times $ t < 1/\omega $. This limitation is expected, given that an ABC aligns more closely to a circle swimmer without rotational noise for larger $ M $.
 
Strikingly, neither the ISF nor MSD fit parameters provide a fully consistent alignment between the ABC propagator and the experimental Hexbug data for short times [see Fig.~\ref{fig:propagator}~a)]. Although the agreement between the experiments and the ABC model generally improves with time, the short-time behavior seems to be qualitatively different. This suggests that subtle deviations in the short-time and small-displacement dynamics exist between the bristle bots and the ABC model --- deviations that are not captured by MSD or ISF analysis alone. These findings point to the need for more refined modeling approaches to accurately describe the propagator of Hexbugs on time scales where the diffusive noise must be taken into account. 

It is important to note that the short-time propagator is highly sensitive to particle displacements between consecutive camera frames. As such, the results may be influenced by tracking artifacts, such as the ``frozen frames'' discussed in Sec.~\ref{sec:experiments}, and a comprehensive analysis of these deviations lies beyond the scope of the present study. While a higher-resolution assessment of particle displacements would be necessary to draw definitive conclusions regarding the discrepancies between the Hexbug propagator and the ABC model, several physical mechanisms may also contribute to the observed deviations.

First, despite the high friction between the MDF surface and the Hexbug, residual inertial effects may persist, which are not accounted for in the ABC framework. Secondly, if inertial contributions are indeed present, the neglected self-alignment terms~\cite{Dauchot2019, Carrillo2025} could influence short-range dynamics. Also, the coupling between the self-alignment terms and translational noise remains unclear and may play a role in shaping the propagator for non-negligible positional fluctuations even in the absence of inertia or confinement. Moreover, our assumption of isotropic translational noise throughout this work may be an oversimplification. Since the width of the propagator directly reflects the translational fluctuations, any anisotropy in positional noise could lead to deviations from the predictions of the isotropic model, which average out in the MSD and the ISF. Finally, it is worth investigating whether such deviations stem from differences between hydrodynamic Stokes friction, which is central to Langevin-type models, and Coulomb contact friction, which dominates the Hexbug-MDF interaction. While both lead to an overdamped motion, the former involves continuous momentum transfer via fluid collisions, whereas the latter dissipates energy through surface impacts while vibrating.

The most conservative explanation for the observed deviation in macroscopic bristle bots is the contribution of inertia. To assess its potential impact on the dynamics, we adopted the underdamped Brownian circle‑swimmer model from Refs.~\cite{Loewen2020, Scholz2018, Sprenger2023} and performed a qualitative test: we numerically integrated the corresponding equations of motion using a stochastic Euler scheme and examined how the short‑time propagator changes for different relaxation times of the linear and angular momenta. Our simulations show that relaxation times on the order of the time lag between two consecutive frames of the video tracking can indeed produce a narrowing of the averaged propagator at short times, consistent with our observations (see Appendix~\ref{app:Inertia} for more details). Nevertheless, drawing firm conclusions requires more systematic investigations, which lie beyond the scope of the present video tracking resolution.

Note also that the determined values for $ D $ are the most volatile parameters with respect to the MSD and ISF fits: Changing the interval of the fits slightly, significantly affects the estimations for the translational noise, where we found deviations of more than a factor of two or three. Such variations of $ D $ depending on the fit procedure are expected as neither the MSD nor the ISF data expresses distinct features reflecting the translational noise (which highlights the usefulness of the propagator). By varying the details of the fit procedure, the translation diffusion constant can be tuned to better align with the propagator results for small times. However, then, significant deviations arise for $ \omega t = 1 $. Hence, the widths of the experimental propagator curves cannot be described by a single constant diffusion coefficient when assuming the ABC framework.

Finally, we want to highlight here that our measurements in the short-time regime resolve the width of the distribution $ \mathbb{P}(r,t) $, which is directly linked to the translational fluctuations. This stands in stark contrast to the evaluation of the MSD and ISF, where the effects of translational noise seem negligible. This observation underscores the high sensitivity of the propagator, making it a particularly valuable tool for evaluating translational noise in macroscopic active matter systems. Translational noise is often overlooked in experiments involving bristle bots, but our findings suggest that $ \mathbb{P}(r,t) $ provides a robust framework for studying it.

\section{Outlook}
\label{sec:discussion}

Recently, bristle bots have gained significant attention in active matter physics as a convenient and accessible platform for studying active agents at the macroscopic scale. These systems offer a unique opportunity to explore active matter phenomena in a controlled and highly visual environment. While additional effects, such as self-alignment under external potentials and confinement, have been investigated, the fundamental question of which physical model best provides a robust foundation for the equations of motion of free Hexbugs remained underexplored. In this work, we addressed this question by investigating whether the ABC model can approximate the dynamics of a free chiral bristle bot. By combining experimental data and analytical predictions, we systematically evaluated its validity across multiple dynamical observables.

Our findings demonstrate that the overdamped equations of motion of the ABC model provide a good approximation of Hexbug dynamics, particularly when analyzed via MSD and ISF. For systems with low rotational diffusion and dominant circular motion, the MSD is a reliable tool for parameter extraction. In contrast, the ISF can offer slightly better accuracy for more diffusive agents on small spatial scales. Deviations in the short-time real-space propagator suggest limitations in the ABC model’s treatment of translational fluctuations (in particular, by a single time-independent diffusion coefficient) and short-time dynamics, though these may stem from video tracking resolution constraints and residual inertia contributions. A deeper investigation into these deviations could further validate or challenge the ABC model’s role in describing the dynamics of Hexbugs.

Beyond modeling, our results highlight Hexbugs as effective macroscopic realizations of ideal ABCs, especially for MSD and ISF studies. This makes bristle bots a widely available, cost-effective, and easy-to-analyze alternative to anisotropic active colloids on the microscale~\cite{Kruemmel2013, Hagen2014}, which are commonly used to investigate chiral active motion. For example, phenomena such as the gravitaxis of active L-shaped particles~\cite{Rusch2024, Hagen2014} could potentially be explored using chiral Hexbugs under the influence of gravity (taking self-alignment into account). The macroscopic scale of Hexbugs offers a convenient platform for testing such phenomena in a controlled and highly visual environment. Investigating these effects in future studies could provide valuable insights into the dynamics of chiral active systems and further establish Hexbugs as a versatile tool for active matter research.

Finally, by applying and comparing three observables (MSD, ISF, and the propagator), we offer a scalable systematic protocol for evaluating dynamical models of macroscopic active agents. Each observable provides a different resolution level, allowing researchers to tailor their analysis to the system’s complexity. We expect this approach to be broadly applicable to bristle-bot variants and other toy models, laying the groundwork for more refined and predictive modeling of macroscopic active agents.\nocite{Data}

\begin{acknowledgments}

The authors thank R. Rusch for the software provided to fit and plot the theoretical prediction of the intermediate scattering function~\cite{Rusch2024_Repo} and S. Anand for valuable discussions and feedback. This research was funded in part by the Austrian Science Fund (FWF) 10.55776/P35580. Y.R. and A.A. acknowledge support from the European Research Council (ERC) under the European Union’s Horizon 2020 research and innovation program (Grant Agreement No. 101002392). Y.R. acknowledge support from the Israel Science Foundation (grant No. 385/21). We acknowledge the use of AI tools to polish, condense, and enhance our writing. 

\end{acknowledgments}

\appendix

\section{Rough approximation of the experimental uncertainties to establish fit weights} \label{app:ERROR}

To approximate the experimental uncertainties, used primarily as weights in the data fitting, we calculate a variance‑inflated standard error of the mean: Since all possible combinations $ \vec{R}_k - \vec{R}_l $ for a given $ t = (k-l) \Delta t $ are used as the sample set, many overlapping ``trajectory pieces'' are included. This introduces correlations between observables calculated from different trajectory pieces, particularly for large $ k-l $. For example, if the total trajectory length is $T=4\Delta t$ and $t = 3\Delta t$, the sample size is $N_\textrm{sample}(t) = 2 $, as there are two possible trajectory pieces with a lag time of $3\Delta t$. However, these two pieces overlap for $66.6\%$ of their length, making them highly correlated. Additionally, adjacent pieces are further correlated by the persistence of the active motion and we cannot fully treat them as independent samples. To address this issue, we approximate an effective sample size $ N_{\textrm{eff}} $ that roughly corrects for the correlations between the trajectory pieces (i.\@e.\@, $ N_{\textrm{eff}} $ counts the effective number of statistically independent trajectory pieces). We motivate this effective sample size by the standard variance-inflation relation~\cite{brooks2011handbook}
\begin{align}
    N_\text{eff}(t) = \frac{N_\text{sample}(t)}{1+2\tau_\text{corr}}, \label{eq:SAMPLE}
\end{align}
which represents a correction to the number of independent samples. Here, $\tau_\text{corr} = \int_0^{\infty}\rho(s)\, \textrm{d}s / \Delta t $ is the integral of the normalized correlation function $ \rho(t) $ of the overlapping trajectory pieces. 

Focusing first on the physical correlations due to the active motion of an ABP, $ \tau_{\textrm{corr}} $ is given by the integral over an exponential decay governed by the persistence time, and, thus, $ \tau_{\textrm{corr}} \propto 1/D_{\textrm{rot}} $. The geometric correlations due to the large overlap of the trajectory pieces are roughly estimated to lead to contributions of $ \tau_{\textrm{corr}} \propto t $. To take both effects into account, we approximate the ``inflation factor of the variance'' (i.\@e.\@, the denominator) via ${1+2\tau_\text{corr} \approx (2/D_{\textrm{rot}}+t)/\Delta t}$, which provides a practical and sufficiently accurate heuristic estimate for our purposes.

For small lag times $1/D_{\textrm{rot}} \gg t$, the correlations between trajectory pieces are mainly determined by the persistence time of the active motion. In this regime, our estimate for the effective sample size becomes $ N_{\textrm{eff}} \approx T D_{\textrm{rot}}/2, $ while the total number of trajectory pieces is $ 
N_{\textrm{sample}} \approx T/\Delta t. $
Intuitively, this means that only one displacement within each time interval of two persistence times provides an independent sample. For large lag times $t \gg 1/D_{\textrm{rot}}$, the expression for the effective sample size simplifies to $N_{\textrm{eff}} \approx T/t, $ which corresponds to counting the number of fully non-overlapping pieces of duration $t$ within the total observation time. Our estimate for the inflation factor can be viewed as a simple interpolation between these two limiting cases.

Note that, while the employed approximation yields physically reasonable results in the limiting cases of small and large lag times, the approximation is expected to be least accurate at intermediate lag times. Moreover, our estimates most accurately reflect the effective sample size of trajectory pieces of an ABP without chiral bias. In our system, the inflation factor should, in principle, exhibit an oscillatory behavior due to the circular motion of the Hexbugs. However, since our estimates are already rather conservative and sufficient for our analysis, we neglect these effects. Finally, the persistence time $1/D_{\textrm{rot}}$ is not known at this stage, as we aim to infer this parameter through weighted fits that incorporate the effective sample size in their weights. 

For the purpose of computing the weights, we use an initial guess of $ 2/D_{\textrm{rot}}\approx 10\, \mathrm{s} $, consistent with the order of magnitude of the persistence reported in previous Hexbug studies~\cite{Altshuler2024, Carrillo2025, Callegari2023}. This value is fixed for all effective sample sizes throughout this paper and should be interpreted as a fixed baseline to mitigate correlations due to persistent motion on the timescale of seconds that is added to the influence of the overlapping trajectory pieces. We verified that varying this baseline does not significantly affect the fit parameters obtained from weighted MSD fits. For instance, for the diffusive Hexbug, the resulting deviations are of the order of $v$: $0.2\%$, $\omega$: $0.5\%$, $D_{\textrm{rot}}$: $5.8
\%$, and $D$: $4.7 \%$ when using $ 2/D_{\textrm{rot}} = 0 $ instead. In general, the translational diffusion coefficient is expected to show a more pronounced variation, as this parameter is intrinsically difficult to resolve from the MSD and ISF data of the Hexbugs. If other physical contributions that lead to additional persistence (such as self-alignment or inertia) are introduced to the dynamics, this baseline can be appropriately adjusted.

The estimated effective sample size is used to approximate the error bars for the MSD shown in Fig.~\ref{fig:msd_fit}. In detail, the uncertainties in this figure represent
\begin{align}
    &\delta\langle\Delta\vec{R}^2(t)\rangle \nonumber \\ &\approx \frac{1}{\sqrt{N_{\textrm{eff}}}}\sqrt{\frac{\sum_{k-l = t / \Delta t} [\langle\Delta\vec{R}^2(t)\rangle -  |\vec{R}_k - \vec{R}_{l}|^2]^2}{N_{\textrm{sample}}}}.
\end{align}
It is important to note that this choice of error bars might overestimate the true experimental uncertainties, as our approximation of the effective sample size is likely to inflate the correlation length of the squared displacements. A more rigorous derivation of the uncertainties is beyond the scope of this work, as we focus on the qualitative correspondence between bristle bots and ABCs. Note, however, that our estimate scales for large lag times $t \gg 1/D_{\textrm{rot}}$ in the same way as the rigorous result for passive particles derived in Ref.~\cite{Qian1991}.

For the ISF fit, we use an equivalent equation for the error bars, using the same estimate of $ N_{\textrm{eff}} $ in the calculation of the uncertainties. This is justified as $ N_{\textrm{eff}} $ approximates the number of displacements $ \Delta \vec{R}(t) $, which can be used for independent estimates of the ISF and the calculation of the corresponding average. However, using the same $ N_{\textrm{eff}} $ for MSD and ISF is an approximation.

Generally, our work does not aim at a precise or exhaustive quantification of the uncertainties of the analyzed observables. The heuristic estimates we apply are fully sufficient for their intended purpose --- namely, to provide reasonable weights for the fits and to demonstrate the excellent agreement between the ABC model and the experimental data. If a more detailed error analysis is required, we refer the reader to Ref.~\cite{Bailey2022}. In particular, to obtain more accurate uncertainty estimates, separate values of $N_{\textrm{eff}}$ for the MSD and ISF must be determined by computing the corresponding covariances independently [e.\@g.\@, inserting the corresponding autocorrelation functions of the observables into Eq.~\eqref{eq:SAMPLE}]. Other recommended options are standard methods such as block averaging or fitting to bootstrapped data sets~\cite{Bailey2022, Fogelmark2018}.

\section{Influence of inertia on the propagator} \label{app:Inertia}

\begin{figure}[ht!] 
    \centering
    \includegraphics[width=\linewidth, trim=0 0 0 0, clip]{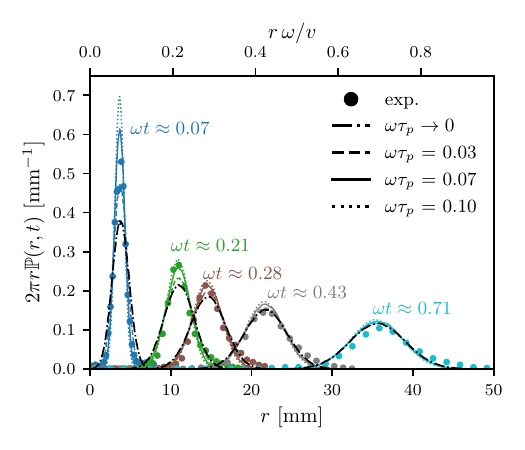}
    \caption{%
Probability distribution $2\pi r\, \mathbb{P}(r,t)$ of the displacement $r$ from the initial position at time $t$ for different momentum relaxation times $\tau_p$. The circles show the experimental data for the Hexbug with the stronger rotational bias. The black lines correspond to $\omega \tau_p \to 0$, which is the active Brownian circle swimmer limit. The color-coded lines are the results from numerically integrating the equations of motion that include inertia contributions. The frame rate of the video tracking is $\omega \Delta t = 0.07$.
}
    \label{fig:propagator_app}
\end{figure}

To assess whether inertial effects could account for the deviations observed in the averaged short-time propagator, we numerically integrate the equations of motion of an underdamped Brownian circle swimmer 
\begin{align}
      \tau_p \,\ddot{\vec{R}} &= - \dot{\vec{R}} + v\,\vec{u} + \sqrt{2D}\,\vec{\eta}(t), \\
    \tau_L \,\ddot{\vartheta} &= - \dot{\vartheta} + \omega + \sqrt{2D_{\textrm{rot}}}\,\xi(t).
\end{align}
as established in Refs.~\cite{Loewen2020, Scholz2018, Sprenger2023} using a stochastic Euler scheme. The parameters $\tau_p$ and $\tau_L$ denote the relaxation times of the linear and angular momentum, respectively. By varying these timescales, we can interpolate between the overdamped regime ($\tau_p \ll \Delta t$ and $ \tau_L \ll \Delta t $) and the strongly underdamped regime ($\tau_p \gg \Delta t$ and $ \tau_L \gg \Delta t $). For simplicity, we set $\tau_p = \tau_L$ throughout the following qualitative analysis.

For each value of $\tau_p$, we run a single simulation of total duration about $ 2080\,\textrm{s} $ with a time step $\delta t = 0.83 \, \upmu \textrm{s}$ of the Hexbug with a stronger rotational bias~\cite{Git}. From the resulting trajectories, we compute the propagator in the same way as for the experimental data, ensuring that the initial velocity has already relaxed before sampling. The results are shown in Fig.~\ref{fig:propagator_app}. We find that inertial contributions can indeed account for the reduced width of the short-time propagator observed for the Hexbugs: when the momentum relaxation time $ \tau_p $ becomes comparable to the frame rate $ \Delta t $, the propagator narrows at the short lag times.

However, because the video frame rate limits our ability to resolve very short timescales, we cannot rule out additional effects that may influence the propagator. In particular, previous studies~\cite{Dauchot2019, Carrillo2025} indicate that when inertia becomes relevant, self‑alignment terms naturally emerge in the effective equations of motion, which could further affect the propagator. A detailed investigation of inertial effects lies beyond the scope of this work, which focuses on the overdamped regime. To confirm or exclude inertia as the primary source of the observed deviations, future experiments with higher frame rates and rotationally resolved dynamics, which also study the long-time limit of the propagator, would be valuable.

\bibliographystyle{apsrev4-1-title}

\end{document}